\documentclass[aps,pra,twocolumn,floatfix,showpacs]{revtex4-1}
\usepackage{amsmath,bm}
\usepackage{graphicx,amsfonts}
\usepackage{latexsym, graphicx, pstricks,rotating, psfrag}
\usepackage[pdftex,bookmarks,colorlinks]{hyperref}
\definecolor{darkblue}{rgb}{0.0,0.0,0.3}
\hypersetup{colorlinks=true,citecolor=darkblue,urlcolor=black, 
unicode=true,pdftitle=draft.pdf,pdfauthor=Ritabrata Sengupta,bookmarks=false}
\begin{document}
\title{Extremal extensions of entanglement
witnesses : Unearthing new bound entangled states}
\author{R. Sengupta}
\email{ph09028@iisermohali.ac.in}
\affiliation{Indian Institute of Science Education \& Research
(IISER) Mohali, Sector 81 Mohali, 140 306 India}
\author{Arvind}
\email{arvind@iisermohali.ac.in}
\affiliation{Indian Institute of Science Education \&
Research (IISER) Mohali, 
Sector 81 Mohali, 140 306 India}
\begin{abstract}
In this paper, we discuss  extremal extensions of 
entanglement witnesses based on Choi's map. The
constructions are based on a generalization of the Choi map
due to Osaka, from which we construct entanglement
witnesses. These extremal extensions are  powerful in
terms of their capacity to detect entanglement of 
positive under partial transpose (PPT)
entangled states and lead to unearthing of entanglement 
of new PPT states. We also use the Cholesky-like
decomposition to construct entangled states which are
revealed by these extremal entanglement witnesses.
\end{abstract}
\maketitle
\section{Introduction}
\label{introduction}
Quantum entanglement plays a central role in quantum theory
from a conceptual as well as a practical point of view.  On
the conceptual front, entanglement is intimately connected
with the notions of non-locality and violation of Bell's
inequalities, which lie at the heart of the way the quantum
mechanical description of the world differs from the
classical one. On the practical front, quantum entanglement
is essential in providing a computational advantage to
quantum computers over their classical
counterparts~\cite{NC}.

The quantum state of a physical system with a 
finite-dimensional complex Hilbert space $H$ 
(which is either pure or mixed),
is represented by $\rho\in\mathcal{B}(H)$ which is a
positive semi-definite hermitian operator with unit trace.
The set of states forms a convex set and the extremal points
of this set are pure states which are operators of rank 1.

For composite quantum systems, the Hilbert space is the
tensor product of Hilbert spaces of the individual systems.
Consider a bipartite quantum system  with its state space
given by $H_A\otimes H_B$ where $H_A$ and $H_B$ are the Hilbert
spaces of individual quantum system.
A bipartite state
$\rho_{AB}\in\mathcal{B}(H_A\otimes H_B)$ is said to be
separable if it is possible to decompose it as follows:
\begin{equation}
\rho_{AB}=\sum_{j=1}^m
p_j\rho_{A_j}\otimes \rho_{B_j};\quad \forall j~~ p_j>0,\quad \sum_{j=1}^m
p_j=1;
\label{separable}
\end{equation}
where $\rho_{A_j}$ and $\rho_{B_j}$ are
states in systems $A$ and $B$ respectively.  
A bipartite state
$\rho\in\mathcal{B}(H_A\otimes H_B)$ is said to be entangled
if it is not separable i.e. it  cannot be  
expanded in the form given in Equation~(\ref{separable}).

For pure quantum states, the characterization of states as
separable or entangled, is easily achieved by computing the
entropy of the reduced density operator of the subsystems.
However, such a characterization for mixed states is a
non-trivial problem.  While a large volume of work
has appeared on this issue in the past two decades, the
problem still remains open. In the absence of a ``final
solution'', explorations into finding new classes of  mixed
entangled states and new ways of constructing them, is
useful and provides insights into the classification
problem.

An important method to detect the entanglement of quantum
states is to construct entanglement witnesses, which are
positive maps that are not completely positive \cite{H1}.  Consider a
positive map $h_A$ defined on the Hilbert space $H_A$ of system
$A$.  If this map is not completely positive, then for some
$H_B$ the map $h_A\otimes I$ acting on $H_A\otimes H_B$ will
not be positive. Therefore, there will be a state
$\rho_{AB}$ for which $ (h_A\otimes I) \rho_{AB}$ will be
negative.  However, this cannot happen for a separable state
defined in~(\ref{separable}).  Therefore, such a
$\rho_{AB}$ has to be entangled.

Partial transposition was the first such entanglement
witness which was used to unearth the entanglement of pure
as well as mixed states \cite{P1,H1}. While negativity under
partial transpose (NPT) indicates entanglement, positivity
under partial transpose is both necessary and sufficient
only for $2\otimes 2$ and $2 \otimes 3$ systems.  The proof
relied on the earlier works of Woronowicz~\cite{wor},
Arvison~\cite{arve69, arve74} and St{\o}rmer~\cite{sto1,
sto2} where they show that in dimension 2, all positive but
not completely positive maps are decomposable 
(can be written as a combination of a completely positive 
map and a transposed completely positive map). 
A corollary of their result shows that, for
dimensions $2\otimes 2$ and $2\otimes 3$, a state is
separable if and only if it remains positive under partial
transpose.  However in higher dimensions there could be maps
which are positive but not completely positive and which may
not have a connection with partial transpose.  This means
that there could be states which are positive under partial
transpose (PPT) and are still entangled.  Such states are
called PPT entangled states.  Obviously for such states,
there exists a positive but not completely positive map
revealing their
entanglement~\cite{jam,wor,wor1,RevModPhys.81.865,guth}.

Choi provided the first example of an indecomposable map
which later led to the detection of entangled states beyond
partial transpose~\cite{choi2}. A number of other discrete
examples are also available \cite{cho1, ha1, os, ha2, ha1, T1, rob1, rob2, rob3, rob4}. However there is no systematic
way of characterizing such maps beyond two dimensions.
A few methods have been recently suggested for generating
examples of positive maps which are not completely positive
\cite{kossak1, kossak2}. However, there is no straight
forward
procedure to verify if they are indecomposable or not.
Generalizations of Choi's method have also
been adopted to produce indecomposable maps which in
principle have the potential to reveal entanglement of new
quantum states~\cite{kossak4}.

The present paper is an attempt in this direction, where we
have constructed extremal extensions of known entanglement
witnesses in order to unearth new classes of bound entangled
states.  We have constructed examples of PPT entangled
states which are revealed by extremal Choi type maps
considered by Choi and Lam \cite{choilam}, and later by
Osaka\cite{os}, and have exploited the
Cholesky decomposition in a nontrivial way. The
Cholesky decomposition has been used by Chru{\'s}ci{\'n}ski
et al. \cite{PhysRevA.77.022113} in the context of PPT
states; however our analysis goes much beyond that.
Moreover, given any extremal positive but not completely
positive map, we can use our method to generate new classes
of extremal maps. In particular, for the $3\otimes3$
bipartite system, we have constructed PPT states whose
entanglement is revealed by our map but is not revealed by
Choi's map.  For the positive maps which are not completely
positive, the question of extremality has been settled only
for a few maps and the problem of determining the structure of such
maps and their extremal or non-extremal nature is in
general difficult.  In this regard, we believe that our
generalization adds new insights into the class of extremal
entangled witnesses.  Further, for a given map which is
positive but not completely positive, it is not always
trivial to find the PPT entangled states revealed by the
map. In our case we were able to find a family of such
states.

The material in this paper 
is arranged as follows: 
In section~\ref{section_2} we discuss the bi-quadratic forms
and their connection with positive maps, where the notions of
extremality and decomposability are discussed as important
ingredients for the purpose of classification of entangled
states. In
section~\ref{section_3} we explore the possibility of
extremal extensions of Choi's map. We
present Osaka's map as an extremal extension of Choi's map
and go on to construct other extremal extensions. We then
turn to constructing examples of bound entangled states whose
entanglement is revealed by Osaka's map and by our new
extremal extension of Choi's map. At the end of this section
we show that the bound entangled states that we have
constructed are robust and can take a certain amount of noise
before they lose their entanglement. 
Section~\ref{section_4} offers some concluding remarks.
\section{Bi-quadratic forms and extremal Maps}
\label{section_2}
Given a non-negative  polynomial
$P(x_1,\cdots,x_n)\geq0,~\forall(x_1,\cdots,x_n)\in\mathbb{R}^n$
of degree $d$, the question  whether $P$ can always be
written as a  sum of squares of polynomials has been around
for a long time.  Minkowski conjectured that in general the
answer should be `no'. It was proved by Hilbert that, except
for three exceptional  cases ($n=1$, $d$ arbitrary; $n$
arbitrary, $d=2$; and one non-trivial case $n=2$, $d=4$),
there always exist positive semi-definite polynomials which
cannot be written as a sum of squares of polynomials.
However, his proof was by an indirect method and did not provide
actual examples of such polynomials.  For a survey and
development of the problem, see Rudin~\cite{rudin}. The generalized
version of this problem on rational polynomials, is known as
Hilbert's 17th problem, and for which the answer is `yes'.

The first counterexample was constructed  by Choi
\cite{choi2} where he considered a positive 
semi-definite
bi-quadratic form $F_{\mu}(X:Y)$ (each term having degree
four), with six variables, divided into two sets denoted by
$X=\{x_1,x_2,x_3\}$, and $Y=\{y_1,y_2,y_3\}$
 given by:
\begin{eqnarray}
F_\mu(X:Y)&=&(x_1^2y_1^2+x_2^2y_2^2+x_3^2y_3^2)
\nonumber\\
&&-2(x_1x_2y_1y_2+x_2x_3y_2y_3+x_3x_1y_3y_1)
\nonumber\\
&&+\mu(x_1^2y_2^2+x_2^2y_3^2+x_3^2y_1^2)
\label{form}
\end{eqnarray}
where $\mu\geq1$.

Choi proved that for $\mu>1$, this bi-quadratic form is
non-negative definite but cannot be written as a sum of
squares of quadratic forms \cite{choi2}. Later it was shown by Choi and
Lam \cite{choilam} that it is also true for $\mu=1$. 
Choi's method has been modified and extended,
and different examples of such positive semi-definite
bi-quadratic forms were found. Among these, the results
by Osaka~\cite{os}, Cho et. al.~\cite{cho1}, and Ha~\cite{ha,
ha1, ha2} are important. Later generalizations of these
methods for generating such forms in arbitrary dimensions
were developed by  Chru{\'s}ci{\'n}ski and
Kossakowski~\cite{kossak4}.
\subsection{Connection with positive maps} 
The intimate
connection between a positive map and positive semi-definite 
bi-quadratic forms was also discovered by Choi
\cite{choi2}. Before describing the connection we 
provide a few definitions.

\noindent{\bf Definition~:~}
A hermiticity preserving map $h:M_n(\mathbb{C})
\longrightarrow M_n(\mathbb{C})$, is said to be a positive
map, if it maps positive semi- definite operators to
positive semi-definite operators.  Here $M_n(\mathbb{C})$
denotes the set of all $n\times n$ complex matrices.

\noindent{\bf Definition~:~}
A positive map $h$ is said to be $k$-positive if the
extended map \[ h\otimes\mathbf{1}_k : M_k(\mathbb{C})\otimes
M_n(\mathbb{C})\longrightarrow  M_k(\mathbb{C})\otimes
M_n(\mathbb{C}),\] is positive, we here $\mathbf{1}_k$ denotes the
identity mapping on the auxiliary space $M_k(\mathbb{C})$.
The map $h$ is said to be completely positive if the above
extensions are positive for all $k\geq1$.

The connection between the maps and bi-quadratic forms can be 
established as follows. Consider a hermiticity preserving linear map  
\begin{equation}
S:\mathbb{C}_m\rightarrow \mathbb{C}_n,
\end{equation}
We can construct the corresponding  bi-quadratic form   $F(X:Y)$  as 
\begin{equation}
F(X:Y)=\langle Y|S(X\cdot
X^T)|Y\rangle
\end{equation}
 where $X=\begin{pmatrix}
x_1\cdots x_m \end{pmatrix}^T$ 
and $Y=\begin{pmatrix}
y_1\cdots y_n\end{pmatrix}^T$. 

On the other hand, let $F(X:Y)$ be  a bi-quadratic form.
Notice that, it is a quadratic form with respect to $Y$ (as
well as $X$). So we can write it in the form $\langle
Y|A_X|Y\rangle$. Thus we get a map which takes any
one-dimensional projection $X.X^T$ to $A_X$. Using linearity
and hermiticity, we can extend it to a map which preserves
hermiticity. It was shown by Choi that, given any positive
semi-definite form, the corresponding map is a positive map
and vice-versa \cite{choi3,choi2}.

There is thus a bijective relation between the set of
positive semi-definite forms and positive maps between
matrix algebras. The property of complete positivity can
also be translated easily. If a map is completely positive,
the corresponding bi-quadratic form can be written as a sum
of squares of quadratic forms and vice versa. Put
differently, if a map is positive but not completely
positive, the corresponding bi-quadratic form will be
positive semi-definite but can not be written as a sum of
squares of quadratic forms.  Thus each such form gives rise
to a unique  map between the space of real symmetric
operators, which  can be trivially extended to the set of
all hermitian operators, and then to all operators. This
also connects with the work of Arvison~\cite{arve69, arve74} and 
St{\o}rmer~\cite{sto1, sto2} who were exploring the set of
positive maps between $C^*$-algebras. Since then, other
examples and classes of such maps have been  discovered.

By the above correspondence, the Choi 
quadratic form given in Equation ~(\ref{form}) leads to the following map for
$3\times 3$ matrices.
\begin{eqnarray}
\Phi_C^I(\mu) : &&\begin{pmatrix} a_{11} & a_{12} &
a_{13}\\ a_{21} & a_{22} & a_{23}\\ a_{31} & a_{32} & a_{33}
\end{pmatrix} \mapsto 
\nonumber \\
&& \begin{pmatrix} a_{11}+\mu
a_{33} & -a_{12} & -a_{13}\\ -a_{21} & a_{22}+\mu a_{11} &
-a_{23}\\ -a_{31} & -a_{32} & a_{33}+\mu a_{22}
\end{pmatrix},
\end{eqnarray}

with $\mu\geq1$. From the quadratic form~(\ref{form}), exchanging 
the $X$ and $Y$ variables, we can get
another map, which is given by 

\begin{eqnarray}
\Phi_C^{II}(\mu) : &&\begin{pmatrix} a_{11} & a_{12} &
a_{13}\\ a_{21} & a_{22} & a_{23}\\ a_{31} & a_{32} & a_{33}
\end{pmatrix} \mapsto 
\nonumber \\
&& \begin{pmatrix} a_{11}+\mu
a_{22} & -a_{12} & -a_{13}\\ -a_{21} & a_{22}+\mu a_{33} &
-a_{23}\\ -a_{31} & -a_{32} & a_{33}+\mu a_{11}
\end{pmatrix}.
\end{eqnarray}

Our interest in these positive but not completely positive
maps is because of their ability to detect entanglement of
quantum states.  For the maps that  
are to be used as entanglement witnesses, 
two notions, namely decomposability and
extremality are very important. We define these
notions below.

\noindent{\bf Definition~:~}
A positive but not completely positive map is
called decomposable, if it can be written as a sum of a
completely positive and a completely co-positive map.

This property was first discussed by Woronowicz \cite{wor}.
Since decomposable maps are obtained by combining a
completely positive map with a transposed completely
positive map, it is clear that they are weaker than partial
transpose in terms of their ability to detect entanglement
and therefore are not of interest.  The interesting point
however is that, given a map which is positive but not
completely positive, there is no standard way to check if it
is decomposable or not!

Since the set of positive maps is a convex set it can be
described by its extremal elements. Therefore, it is most
natural to study  extremal positive maps. From the point
of view of detecting entanglement, any extremal map is more
powerful than the maps which are internal points of the set
of positive maps~\cite{choilam,os}.  Choi and Lam define an
extremal map using the corresponding bi-quadratic
form~\cite{choilam} as follows.

\noindent {\bf Definition~:~}
A positive semi-definite  bi-quadratic form $F$ is said to
be extremal if, for any decomposition of $F=F_1+F_2$ where
$F_i$'s are positive semi-definite bi-quadratic forms, 
$F_i=\lambda_iF$, where $\lambda_1,~\lambda_2$ are
non-negative real numbers with $\lambda_1+\lambda_2=1$.

It was shown by Choi and Lam that in the case  $\mu=1$,  the
form $F_1$ defined in Equation~(\ref{form}) is extremal.

Since the set of
positive semi-definite forms is a convex set, it is enough
to identify the set of such extremal forms. At this stage it
is useful to change the notation to the original Choi-Lam
notation for the bi-quadratic forms 
and we therefore denote $F(X:Y)$ as
$F\begin{pmatrix} x_1 & x_2 & x_3\\ y_1 & y_2 &
y_3\end{pmatrix}$. 
\section{Extremal maps and bound entangled states}
\label{section_3}
Osaka extended the result of Choi and Lam, and
generated a class of extremal maps~\cite{os}. 
Osaka's map $\Phi_O(x,y,z)$ is defined as 
\begin{eqnarray}
\Phi_O(x,y,z) :&&
\begin{pmatrix}
a_{11} & a_{12} & a_{13}\\
a_{21} & a_{22} & a_{23}\\
a_{31} & a_{32} & a_{33}
\end{pmatrix}
\mapsto 
\nonumber \\
&&
\begin{pmatrix}
a_{11}+xa_{33} & -a_{12} & -a_{13}\\
-a_{21} & a_{22}+ya_{11} & -a_{23}\\
-a_{31} & -a_{32} & a_{33}+za_{22}
\end{pmatrix},
\label{osaka_map}
\end{eqnarray}
where $x,y,z>0$ and $xyz=1$. 
Osaka showed that this class of maps is
extremal~\cite{os}.

Generalizing beyond Osaka construction, we define  a class of
extremal bi-quadratic forms as follows:
 
Let $F=F\begin{pmatrix} x_1 & x_2 & x_3\\ y_1 & y_2 &
y_3\end{pmatrix}$
be an extremal positive semi-definite
bi-quadratic form. 

For a set of 
non-zero positive real numbers $a,b,c$ we define
\begin{equation}
G\begin{pmatrix} x_1 & x_2 & x_3\\ y_1 & y_2 &
y_3\end{pmatrix}=F\begin{pmatrix} ax_1 & bx_2 & cx_3\\ y_1 & y_2 &
y_3\end{pmatrix}
\end{equation}

It turns out that the form $G\begin{pmatrix} x_1 & x_2 &
x_3\\ y_1 & y_2 & y_3\end{pmatrix}$ is positive
semi-definite and extremal. 

{\bf Proof~:~} We first prove the positivity.  Let us assume that the 
proposition is not true and there exists real numbers $p_1, p_2, p_3, q_1,
q_2, q_3$ such  that $G\begin{pmatrix} p_1 & p_2 & p_3\\ q_1 & q_2 &
q_3\end{pmatrix}<0$. But then by definition $F\begin{pmatrix} p_1' & p_2' & p_3'\\ q_1 & q_2 &
q_3\end{pmatrix}<0$; for a set of real numbers $p_1', p_2', p_3', q_1,
q_2, q_3$ where $p_1'=ap_1,~p_2'=bp_2$ and $p_3'=cp_3$. This contradicts the 
assumption that $F$ is a positive semi-definite form for all real 
values of $x_i$'s and $y_j$'s.

For extremality, let us assume 
$G=G_1+G_2$, where $G_1$ and $G_2$ are positive semi-definite
bi-quadratic forms. 

Notice that 
\begin{eqnarray*}
F\begin{pmatrix} x_1 & x_2 & x_3\\ y_1 & y_2 &
y_3\end{pmatrix}&=&G\begin{pmatrix} \frac{x_1}{a} &
\frac{x_2}{b} & \frac{x_3}{c}\\ y_1 & y_2 &
y_3\end{pmatrix}\\
&=&G_1\begin{pmatrix} \frac{x_1}{a} &
\frac{x_2}{b} & \frac{x_3}{c}\\ y_1 & y_2 &
y_3\end{pmatrix}+G_2\begin{pmatrix} \frac{x_1}{a} &
\frac{x_2}{b} & \frac{x_3}{c}\\ y_1 & y_2 &
y_3\end{pmatrix}, 
\end{eqnarray*}
as we have assumed. Define two positive semi-definite forms
\[F_i\begin{pmatrix} x_1 & x_2 & x_3\\ y_1 & y_2 &
y_3\end{pmatrix}=G_i\begin{pmatrix} \frac{x_1}{a} &
\frac{x_2}{b} & \frac{x_3}{c}\\ y_1 & y_2 &
y_3\end{pmatrix},\]
for $i=1,2$. We can now write,
\[F=F_1+F_2.\]
However,
 $F$ is extremal. Therefore, $F_i=\lambda_iF,~i=1,~2$, and 
 $\lambda_1$ and $\lambda_2$ are positive real numbers
with $\lambda_1+\lambda_2=1$. Thus $G_i\begin{pmatrix} \frac{x_1}{a} &
\frac{x_2}{b} & \frac{x_3}{c}\\ y_1 & y_2 &
y_3\end{pmatrix}=\lambda_iF\begin{pmatrix} x_1 & x_2 & x_3\\ y_1 & y_2 &
y_3\end{pmatrix}$. Hence
\begin{eqnarray*}
G_i\begin{pmatrix} x_1 & x_2 & x_3\\ y_1 & y_2 &
y_3\end{pmatrix}&=&G_i\begin{pmatrix} a\frac{x_1}{a} &
b\frac{x_2}{b} & c\frac{x_3}{c}\\ y_1 & y_2 &
y_3\end{pmatrix}\\
&=&\lambda_iF\begin{pmatrix} ax_1 & bx_2 & cx_3\\ y_1 & y_2 &
y_3\end{pmatrix}\\
&=&\lambda_iG.
\end{eqnarray*}
Since $\lambda_1,~\lambda_2\geq0$ and $\lambda_1+\lambda_2=1$; the form $G$ is an extremal form.

The maps corresponding to the bi-quadratic form $G$  defined
above are positive maps which are  extremal.  
This construction is extendable to higher dimensions 
without any further work. This means that  
given any extremal positive semi-definite bi-quadratic form, $F=F\begin{pmatrix}
x_1 & x_2&\cdots & x_n\\ y_1 & y_2 &\cdots&
y_n\end{pmatrix}$,  for any non zero positive real
$a_1,a_2,\cdots,a_n$; $G=F\begin{pmatrix} a_1x_1 &
a_2x_2&\cdots & a_nx_n\\ y_1 & y_2 &\cdots&
y_n\end{pmatrix}$ is positive semi-definite and extremal.

We now turn to the map corresponding to the extremal
bi-quadratic form $G$ from Equation~(\ref{form}). After working out the
details, the map turns out to be
\begin{eqnarray}
\Phi(a,b,c) :&&\begin{pmatrix}
x_{11} & x_{12} & x_{13}\\
x_{21} & x_{22} & x_{23}\\
x_{31} & x_{32} & x_{33}
\end{pmatrix}
\mapsto 
\nonumber \\
&&\!\!\!\!\!\!\!\!\!\!\!\!\!\!\!\begin{pmatrix}
a^2 x_{11}+c^2 x_{33} & -abx_{12} & -acx_{13}\\
-abx_{21} & b^2x_{22}+a^2x_{11} & -bcx_{23}\\
-acx_{31} & -bcx_{32} & c^2x_{33}+b^2x_{22}
\end{pmatrix}
\end{eqnarray}
where $a,b,c\neq0$.
The map $\Phi(a,b,c)$ is an extremal positive map which may not be
completely positive. 
For the values of $a,b$ and $c$ for which it is not
completely positive, it can act as an entanglement witness.

We now construct a set of PPT entangled states for which the
above map acts as an entanglement witness.
Consider a density operator for a $3 \otimes 3$ system
defined by two parameters $t$ and $x$.
\begin{equation}\label{choiexample}
\rho(x,t)=\frac{1}{4+\frac{3}{t}+4 t}\left(
\begin{array}{ccc|ccc|ccc}
 1+t & 0 & 0 & 0 & x & 0 & 0 & 0 & x \\
 0 & t & 0 & x & 0 & 0 & 0 & 0 & 0 \\
 0 & 0 & \frac{1}{t} & 0 & 0 & 0 & x & 0 & 0 \\\hline
 0 & x & 0 & \frac{1}{t} & 0 & 0 & 0 & 0 & 0 \\
 x & 0 & 0 & 0 & 1+t & 0 & 0 & 0 & x \\
 0 & 0 & 0 & 0 & 0 & t & 0 & x & 0 \\\hline
 0 & 0 & x & 0 & 0 & 0 & 1 & 0 & 0 \\
 0 & 0 & 0 & 0 & 0 & x & 0 & \frac{1}{t} & 0 \\
 x & 0 & 0 & 0 & x & 0 & 0 & 0 & 1
\end{array}
\right).
\end{equation}
This $\rho$ is a unit trace density operator for $t>0$ and
$0\leq x\leq1$.

The action of the map $\Phi(a,b,c)$ on the density operator
$\rho(x,t)$ leads to the transformed density operator
$\rho^{x,}$.
\begin{equation}
\rho^{\prime}(x,t)=(\Phi(a,b,c)\otimes\mathbf{1}_3) \rho(x,t).
\end{equation}

We compute  eigenvalues of $\rho(x,t)^{\prime}$ and use
the negativity of the least eigenvalue as an indicator of
entanglement of $\rho(x,t)$.
It is useful to note that 
the map $\Phi(a,b,c)$ with, $a=b=c=1$ reduces to  Choi's
map $\Phi_C^I(1)$ while for other values of $a,b$ and $c$
it is still an extremal map with a potential to reveal
entanglement of quantum states.

\begin{figure}[htp]
\psfrag{t}{\large $t$}
\psfrag{x}{\large $x$}
\includegraphics[width=7.5cm]{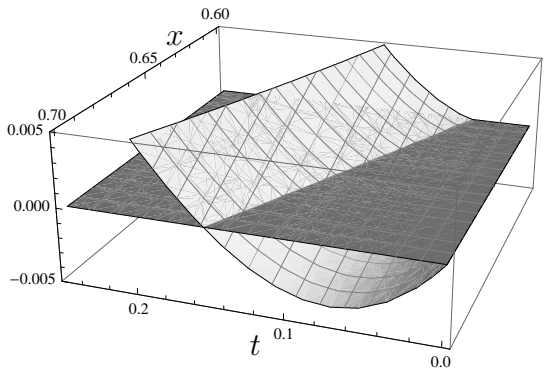}
\caption{
\label{figure1}
The plot of least eigenvalues of $\rho^{\prime}(x,t)$ as
a function of $x$ and $t$. 
The curved surface corresponds to the case where $\rho^{\prime}(x,t)$ was
generated by the action of $\Phi(1+\frac{3}{5},1,1)$ upon
$\rho(x,t)$.  The middle plane is the plane $xy=0$, given a referral plane for highlighting the negativity of the eigenvalues represented by the curved surface.
}
\end{figure}

\begin{figure}[htp]
\includegraphics[width=7.5cm]{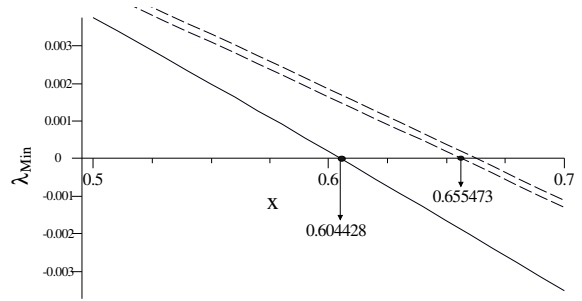}
\caption{
\label{figure2}
The section corresponding to $t=\frac{1}{20}$ of
Figure~\ref{figure1} is displayed here. The blue line
corresponds to the map $\Phi(1+\frac{3}{5},1,1)$ while
the other two curves correspond to the two Choi's maps.  The
window of values of $x$ (approximately $x\geq 0.6025$  and $x<
0.66$) where the map $\Phi(1+\frac{3}{5},1,1)$ is able
to reveal the entanglement of $\rho(x,t)$ and where Choi's
maps do not reveal the entanglement is clearly visible.} 
\end{figure}

A computation of eigenvalues of $\rho^{\prime}(x,t)$ reveals
that for this example, the maps $\Phi(a,b,c)$ have more potential than
Choi's map in unearthing the entanglement of PPT quantum
states.  The results are displayed graphically in
Figures~\ref{figure1} and \ref{figure2}.  We take the
parameter values to be $a=1+\frac{3}{5}$, $b=c=1$ and
calculate the minimum eigenvalues of $\rho(x,t)$ for the
range $x\in[0,1]$ and $t\in(0,1]$.  The results are
displayed in Figure~\ref{figure1}. The curved surface denotes
the minimum eigenvalue of $\rho(x,t)$ after the action
$\Phi(1+\frac{3}{5},1,1)$.   To show the power of this map clearly,
we display a section of the above graph where we fix the
parameter $x=\frac{1}{20}$ and plot the minimum eigenvalue
as a function of $t$. We compare our result with Choi's maps.   
The result is shown in Figure~\ref{figure2}. The continuous line here denotes the minimum
eigenvalue corresponding to  the action
$\Phi(1+\frac{3}{5},1,1)$ while the other two dashed  lines are the
minimum eigenvalues corresponding to $\Phi_C^I$ and
$\Phi_C^{II}$.  It turns out that approximately
after $x=0.604428$, the minimum eigenvalue becomes negative
under $\Phi(1+\frac{3}{5},1,1)$ while the minimum
eigenvalues under $\Phi_C^I(1)$ and $\Phi_C^{II}(1)$ still
remain positive. It is only after $x$ crosses the value
$0.66$ that Choi's maps begin to detect entanglement for
this class of states.  Therefore, for
$\rho(x,\frac{1}{20})$, there is a clear window of $x$
values where the entanglement is revealed by
$\Phi(1+\frac{3}{5},1,1)$ and is not revealed by any of the
Choi's maps. 

The map $\Phi(1+\frac{3}{5},1,1)$ was chosen as a
representative example. In fact the class of maps
$\Phi(a,b,c)$ can reveal the entanglement of a large class of
PPT entangled states and therefore provide a genuine
extremal extension of Choi's maps.

\subsection{PPT entangled states detected by Osaka's map}
We now address the question of constructing PPT
entangled states for which Osaka's map acts as an
entanglement witness.  Although the three-parameter family
of maps due to Osaka as described in
Equation~\ref{osaka_map} has been defined, and is known to
be positive but not completely positive, there has not been
an explicit construction of PPT entangled states whose
entanglement is revealed by this class of maps.

We set up a computer search over the PPT entangled states and
employ the Cholesky decomposition~\cite{bhatia1}  to selectively
scan the states in the Hilbert space. We also make an
intelligent use of Choi-Jamio{\l}kowski isomorphism
developed in~\cite{choi1,choi2,jam,CK} to check at each
stage of our search that the state remains entangled. 
This method of searching for entangled states for a given
positive but not completely positive map is in fact more
general and can be tried for other maps too.

According to the Cholesky decomposition, every density
matrix $\rho$ of a quantum system can  be decomposed as
$\rho=T^\dag T$, where $T$ is an upper triangular
matrix~\cite{bhatia1}.  If $\rho$ is strictly positive (i.e.
all eigenvalues are greater than zero), then the
corresponding Cholesky Decomposition is unique.  For
simplicity, we restrict ourselves to the case of real states
where elements of the density matrix are real and in this
case the Cholesky decomposition reduces to $\rho= T^t
T$, where $t$ denotes the transpose operation.

The Choi Jamio{\l}kowski isomorphism
developed in~\cite{choi1,choi2,jam,CK}  provides a simple
one-way test of entanglement for a given positive map which
is not completely positive. Consider a composite system 
$H_1\otimes H_2$ where both the subsystems are of dimension
$d$. Using the standard basis $\vert i\rangle$ in both the
$H_1$ and $H_2$ we define a maximally entangled state  
\begin{equation}
\vert \psi\rangle=\frac{1}{\sqrt{d}}\sum_{i=0}^{d-1}\vert 
i\rangle\otimes \vert i\rangle;
\end{equation}
Given a positive map  $\Phi$ that is not completely
positive we define an operator
\begin{equation}
W_\Phi=\frac{1}{\sqrt{d}}\sum_{i=0}^{d-1}\vert i\rangle\langle
i\vert \otimes\varphi(\vert i\rangle\langle i\vert). 
\end{equation}
Given a density matrix $\rho$ defined on $H_1\otimes H_2$ 
the operator $W$ provides  a sufficient condition for 
entanglement 
\begin{equation}
{\rm Tr}(W \rho) <  0 \Longrightarrow \quad\rho\quad {\rm Entangled}
\label{choi-jam}
\end{equation}
This is a one-way condition and ${\rm Tr}(W \rho) \geq 0$
does not imply that the state $\rho$ is separable.
The condition for entanglement given in~(\ref{choi-jam}) 
helps us in quickly identifying states whose entanglement is
revealed by the map $\Phi$ and we employ this condition in
our search for PPT entangled states revealed by Osaka family
of maps.

We first
construct a upper triangular matrix $T$. We further restrict
to those states which are invariant under partial transpose
by imposing the condition $(T^t T)^{PT}=T^t T$ (where $PT$
denotes the transpose with respect to the second system).
This amounts to 
generating a  non-trivial  solution of
the equation 
\begin{equation}
(T^t T)^{PT}-T^t T=0.
\end{equation}
It is not always possible to find a non-trivial solution to
the above equation. However, if we begin with a sparse
matrix $T$ we can hope to find a solution to the above
equation.  In our search we also at every stage impose the
following condition
\begin{equation}
\label{sym1} \left\{\begin{array}{l}
\mathrm{Tr}\left(W\Phi_O\rho\right)<0\\
\mathrm{Tr}\left(W\Phi_{C^I}\rho\right)\geq0 \end{array}\right.
\end{equation} 
This means that we restrict ourselves to those PPT states
whose entanglement is revealed by Osaka's map but is not
revealed by Choi's map.
The above methodology guides us in our computer search process
to look for classes of states whose entanglement is revealed
by the Osaka family of maps.

Using this method and employing a computer search protocol,
we construct an example of a PPT
entangled state for a $3\otimes 3$ system. The  upper
triangular matrix $T$ given by
\begin{equation}
\left( \begin{array}{ccccccccc}
\sqrt{10} y & 0 & 0 & 0 & \frac{2+5 y}{\sqrt{10}} & 0 & 0 &
0 & \frac{3 (1+y)}{\sqrt{10}} \\ 0 & y & 0 & 2+5 y & 0 & 0 &
0 & 0 & 0 \\ 0 & 0 & 3 y & 0 & 0 & 0 & 1+y & 0 & 0 \\ 0 & 0
& 0 & 0 & 0 & 0 & 0 & 0 & 0 \\ 0 & 0 & 0 & 0 & \frac{4+5
y}{\sqrt{10}} & 0 & 0 & 0 & \frac{1+y}{\sqrt{10}} \\ 0 & 0 &
0 & 0 & 0 & 1+2 y & 0 & 1+y & 0 \\ 0 & 0 & 0 & 0 & 0 & 0 & 0
& 0 & 0 \\ 0 & 0 & 0 & 0 & 0 & 0 & 0 & 0 & 0 \\ 0 & 0 & 0 &
0 & 0 & 0 & 0 & 0 & 0 \end{array} \right). 
\end{equation}
leads to a one parameter family of density operators,
parameterized by a positive parameter $y$.
\begin{widetext}
\begin{eqnarray}
&&
\rho(y)=\frac{1}{N}\times \nonumber \\
&&\left[\!\!
\begin{array}{ccccccccc}
 10 y^2 & 0 & 0 & 0 & y (2+5 y) & 0 & 0 & 0 & 3 y (1+y) \\
 0 & y^2 & 0 & y (2+5 y) & 0 & 0 & 0 & 0 & 0 \\
 0 & 0 & 9 y^2 & 0 & 0 & 0 & 3 y (1+y) & 0 & 0 \\
 0 & y (2+5 y) & 0 & (2+5 y)^2 & 0 & 0 & 0 & 0 & 0 \\
 y (2+5 y) & 0 & 0 & 0 & 2+6 y+5 y^2 & 0 & 0 & 0 & (1+y) (1+2 y) \\
 0 & 0 & 0 & 0 & 0 & (1+2 y)^2 & 0 & (1+y) (1+2 y) & 0 \\
 0 & 0 & 3 y (1+y) & 0 & 0 & 0 & (1+y)^2 & 0 & 0 \\
 0 & 0 & 0 & 0 & 0 & (1+y) (1+2 y) & 0 & (1+y)^2 & 0 \\
 3 y (1+y) & 0 & 0 & 0 & (1+y) (1+2 y) & 0 & 0 & 0 & (1+y)^2
\end{array}
\!\!\right] \nonumber \\
\label{osexample}
\end{eqnarray}
\end{widetext}
where $N=10+36 y+57 y^2$ is the normalization factor such
that $\mathrm{Tr}(\rho(y))=1$. 
We apply a one parameter subfamily of  Osaka's map defined
in~(\ref{osaka_map}) $\Phi_O\left(1,x,\frac{1}{x}\right)$ to
the family of states $\rho(y)$  and compute the eigenvalues 
of the resultant  operator. We do a similar
computation of the eigenvalues of the operator which is
obtained by the action of Choi's maps $\Phi_C^I$ and 
$\Phi_C^II$ for comparison.

\begin{figure}[htp]
\psfrag{y}{\large $y$}
\psfrag{x}{\large $x$}
\includegraphics[width=7.5cm]{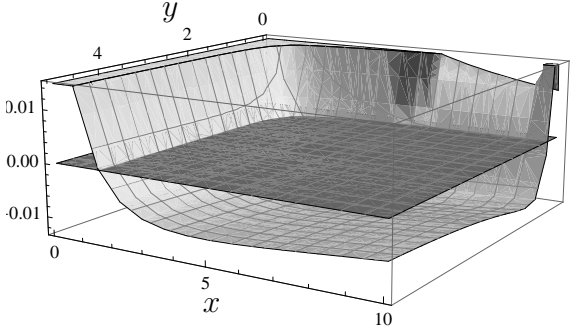}
\caption{\label{figure3}
Application of Osaka's map
$\Phi_O\left(1,x,\frac{1}{x}\right)$ to the density
operator $\rho(y)$. The two independent variables are $x$ and $y$, and
the vertical axis denotes the eigenvalues of $\rho(y)$ under
the map. The curved surface represents
the variation of the minimum eigenvalue. The plane in the
center is the plane $xy=0$, which highlights the portion of the
surface with negative eigenvalue.
}
\vspace{3mm}
\end{figure}
\begin{figure}[htp]
\includegraphics[width=7.5cm]{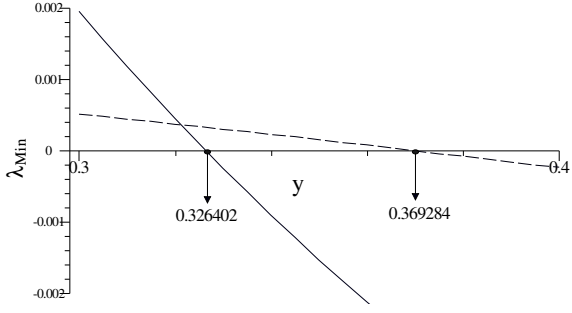}
\caption{\label{figure4}
Application of Osaka's map with $x=6$. The horizontal axis
is $y$ and eigenvalues are along the vertical axis. The continuous line
denotes the variation of eigenvalues under
$\Phi_O\left(1,6,\frac{1}{6}\right)$. The dashed line
corresponds to Choi's map. It clearly shows that
approximately after $y=0.326402$ onward, PPT entanglement is
revealed by $\Phi_O(\left(1,6,\frac{1}{6}\right)$. However
Choi's map can reveal entanglement approximately from
$y=0.369284$. This shows the superiority of Osaka's map
over Choi's map for this instance of $\rho(y)$.  }
\end{figure}
In Figure~\ref{figure3} the least eigenvalue is plotted as a
function $x$ and $y$. Here the curved surface denotes the
minimum eigenvalue corresponding to the
$\Phi_O\left(1,x,\frac{1}{x}\right)$. The middle plane
denotes the $xy$ plane which is placed to indicate the place
when the surface becomes negative. In Figure~\ref{figure4}
we have taken a fixed value of $x$.  The eigenvalues are
plotted along the vertical axis and $y$ varies along the
horizontal axis.  We apply the map
$\Phi_O\left(1,6,\frac{1}{6}\right)$ to this state $\rho(y)$
and plot the minimum eigenvalue which is denoted by the
continuous curve in Figure~\ref{figure4}. The dashed curve
denotes the minimum eigenvalue achieved by the Choi's map.
The plot highlights that approximately after point
$y=0.326402$ the minimum eigenvalue under
$\Phi_O\left(1,6,\frac{1}{6}\right)$ becomes negative. Thus
there is a range of values, where Osaka's map can identify
more PPT entangled states while the Choi's map fails to do
so.

\subsection{Robustness analysis}
\par 
We now consider the robustness of the states given
in Equations~(\ref{osexample}) and~(\ref{choiexample}).  Let $\rho$ be an
arbitrary entangled state. We consider the convex
combination of $\rho$ with a maximally mixed state. For this
case we consider the following convex combination;
\[\rho'(\varepsilon)=\frac{\varepsilon}{9}\mathbb{I}_9+(1-\varepsilon)\rho;\]
and try to detect the range of $\varepsilon$ for which $\rho'$ is
entangled. $\mathbb{I}_9$ denotes the identity matrix of
dimension 9.  Typically, the map which detects entanglement
of $\rho$ is used on $\rho'$ as well. 

\par
\begin{figure}[h]
\includegraphics[width=7.5cm]{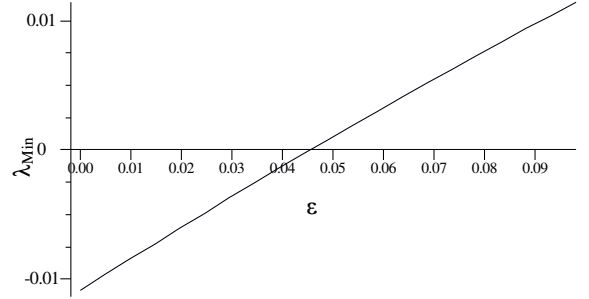}
\caption{\label{figure5}
Application of Osaka's map $\Phi_O$ on the convex
combination $\rho'\left(\varepsilon,
\frac{5}{2}\right)=\frac{\varepsilon}{9}\mathbb{I}_9+(1-\varepsilon)\rho\left(\frac{5}{2}\right)$.
$\varepsilon$ is plotted along the horizontal axis, and eigenvalues
along the vertical axis.  The line shows the
minimum eigenvalue as a function of the robustness parameter
$\varepsilon$. The state remains entangled
approximately up to $\varepsilon=0.045$.}
\end{figure}

 
We begin with $\rho(y)$ of the example~(\ref{osexample}).
Using the process previously described, we  obtain the new state
\[\rho'(\varepsilon,y)=\frac{\varepsilon}{9}\mathbb{I}_9+(1-\varepsilon)\rho(y).\]
For $y=\frac{5}{2}$, ~ $\rho(y)$ is an entangled state, whose
entanglement is revealed by $\Phi_O$. We use the map
$\Phi_O$ on  the family of states $\rho'\left(\varepsilon,
\frac{5}{2}\right)$ and  can see that there is a continuous
range  of $\varepsilon$ for which
$\rho'\left(\varepsilon,\frac{5}{2}\right)$ remains entangled.  The
change in eigenvalues is shown in Figure~\ref{figure5}. It
shows that approximately up to $\varepsilon=0.047$, the state
remains entangled. 

\par
\begin{figure}[h]
\vspace*{24pt}
\includegraphics[width=7.5cm]{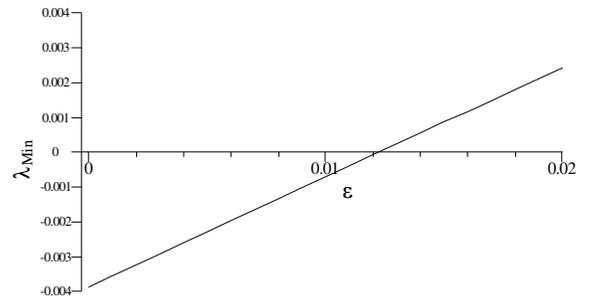}
\caption{\label{figure6}
Application of  $\Phi\left(1+\frac{6}{10},1,1\right)$ on
$\rho'\left(\varepsilon,\frac{7}{10},\frac{3}{40}\right)$. $\varepsilon$
is plotted along the horizontal axis, and eigenvalues along
the vertical axis.  The line shows the 
minimum eigenvalue as a function of the robustness parameter
$\varepsilon$. The state remains entangled
upto approximately $\varepsilon=0.012$.}
\end{figure}

 
We now use the same procedure for $\rho(x,t)$ of the example
in~(\ref{choiexample}). The family of states is given by;
\[\rho'(\varepsilon,x,t)=\frac{\varepsilon}{9}\mathbb{I}_9+(1-\varepsilon)\rho(x,t).\]
We use $\rho\left(\frac{7}{10},\frac{3}{40}\right)$, which
is an entangled state and its entanglement is revealed by
$\Phi\left(1+\frac{6}{10},1,1\right)$.  Now the family
$\rho'\left(\varepsilon,\frac{7}{10},\frac{3}{40}\right)$ is
dependent on $\varepsilon$. We  plot the minimum
eigenvalue of this family as a function of the robustness
parameter $\varepsilon$.  Figure \ref{figure6}
shows that up to approximately $\varepsilon=0.012$ the state
remains entangled.

\section{Conclusions}
\label{section_4}
This work is an exploration in the context of finding
bound entangled states whose entanglement is revealed by
witnesses based on positive maps that are not completely
positive. We have managed to extend the construction to a class of
bound entangled states using the Cholesky decomposition whose
entanglement is revealed by Osaka's map acting as a witness.
Furthermore, we have generated a family of extremal extensions of
Choi's original map and shown that these extremal
extensions are capable of revealing the entanglement of new
classes of entangled states. We are extending this work to 
develop a more general framework where extremal movement
in the map space is tracked down to a similar movement in
the space of bound entangled states. 
Those results will be
presented elsewhere.
\end{document}